\documentclass[12pt]{article}
\usepackage[page]{appendix}
\usepackage{rotating}
\usepackage{graphicx}
\usepackage{longtable}
\graphicspath{{figures/}} 
\usepackage{caption}
\usepackage{subcaption}
\usepackage{epsfig}
\usepackage[sort, round, authoryear]{natbib}
\usepackage{multirow}
\usepackage{algorithm}
\usepackage{algorithmic}
\usepackage{booktabs}
\usepackage{epstopdf}
\usepackage{comment}
\usepackage{mathrsfs}
\usepackage{color}
\usepackage{float}
\usepackage{rotating}
\usepackage{threeparttable}
\usepackage{longtable} 
\usepackage{float}

\usepackage{natbib}
\usepackage{amsmath}
\usepackage{color}

\usepackage[utf8x]{inputenc} 

\usepackage{amsmath, amsthm, amssymb}
\usepackage[bottom]{footmisc}

\begin{document}
\begin{center}
{\LARGE Assessing Intervention Strategies for Non Homogeneous Populations Using a Closed Form Formula for $R_0$} \\ [12pt]
\end{center}
\medskip
\begin{center}
  {\large
  \mbox{Zeynep G{\"o}k{\c c}e {\.I}{\c s}lier} $\bullet$
    \mbox{Wolfgang H{\"o}rmann} $\bullet$
    \mbox{Refik G{\"u}ll{\"u}} 
      }\\[12pt]
  {\it
    Bogazici University
    Industrial Engineering Department,
    Bebek 34342 Istanbul Turkey \\[3pt]
    \mbox{zeynep.yildiz@boun.edu.tr} $\bullet$
    \mbox{hormannw@boun.edu.tr} $\bullet$
    \mbox{refik.gullu@boun.edu.tr} 

  }
\end{center}

\noindent \hrulefill

\noindent {\bf Abstract} \\ A general stochastic model for susceptible $\rightarrow$ infective $\rightarrow$ recovered (SIR) epidemics in non homogeneous populations is considered. The heterogeneity is a very important aspect here since it allows more realistic but also more complex models. The basic reproduction number $R_0$, an indication of the probability of an outbreak for homogeneous populations does not indicate the probability of an outbreak for non homogeneous models anymore, because it changes with the initially infected case. Therefore, we use ``individual $R_0$'' that is the expected number of secondary cases for a given initially infected individual. Thus, the effectiveness of intervention strategies can be assessed by their capability to reduce individual $R_0$ values. Also an intelligent vaccination plan for fully heterogeneous populations is proposed. It is based on the recursive calculation of individual $R_0$ values.  


\baselineskip 20pt plus .3pt minus .1pt \noindent\hrulefill
\section{Introduction}

Epidemiological information is mostly used to plan and evaluate strategies that prevent disease spread by identifying risk factors. Therefore, various disease spread models were developed in the literature. This paper considers a stochastic model for susceptible $\rightarrow$ infective $\rightarrow$ recovered (SIR) epidemics among non homogeneous populations. The basic reproduction number $R_0$ as the expected number of secondary cases produced by a single infected case in a totally susceptible population is important for the determination of the outbreak probability under homogeneous mixing assumption \citep{linda2000comparison, craft2013estimating, hernandez2002markov, kumar2019deterministic}. In order to assess the dynamics of disease behavior, \citet{ross2011invasion} analyses the effect of population size and disease time distribution on $R_0$. Moreover, heterogeneity in population has important effects on disease spread behavior as discussed eg. in \citep{inaba2012new, meyers2007contact} with its realistic assumptions.

This paper is concerned with the notion of computable $R_0$ for heterogeneous models. Our aim is to calculate the expected number of secondary cases produced by a unique given infected case and use it to develop effective intervention strategies and assess the intervention strategies without simulation. \cite{watson1980useful} proposes a model with two mixing levels and defines three types of outbreaks: localized, restricted and generalized outbreaks. There are also others considering $R_0$ for non homogeneous populations to estimate outbreak probabilities. Since the definition of $R_0$ for heterogeneous populations is complex, \cite{diekmann1990definition} and \cite{trapman2007analytical} estimates bounds for the expected infectivity based on disease length and heterogeneity state regarding deterministic disease spread models among totally heterogeneous populations. Because the analysis of stochastic disease spread models for heterogeneous populations is difficult, either simulation is used for the analysis \citep{longini2005containing, ajelli2010comparing} or the models are simplified by decreasing the number of mixing levels. There are a number of recent papers on epidemics in heterogeneous populations in order to estimate $R_0$. They implement agent based simulation claiming that $R_0$ cannot be directly calculated \citep{longini2004, lipsitch2003transmission} while \citet{ball1986unified} calculate the average $R_0$ exactly among a stratified population with two levels by applying branching process methods. \citet{keegan2016estimating} estimate $R_0$ for finite populations with different heterogeneity types to analyse the effects of heterogeneity on basic reproduction numbers. However, a single $R_0$ value is calculated in all of these models to analyse the possibility of an outbreak. \citet{artalejo2013exact} measures $R_{e,0}$ that is the exact number of secondary cases generated by the tagged infected individual if the epidemic starts or is already in progress. They emphasize the probability distribution of the number of secondary cases rather than  its mean. Following \citet{artalejo2013exact}, \citet{economou2015stochastic} determine the distribution of the number of secondary cases for SIS models with exponential disease time. \citet{lopez2016stochastic} introduces $R_x^{exact}$ as the random variable denoting the number of individuals directly infected by a given infected individual during his infectious period given the current state of the process $x$ and its probability mass function is investigated.

Moreover, $R_0$ is generally used in the literature for analyzing the possibility of an outbreak even if it is also possible to use it for intervention strategy analysis. In the literature, to develop and analyze epidemic control strategies, different mathematical approaches are implemented like introducing contact network epidemiology \citep{dimitrov2010mathematical}, implementing optimal control tools \citep{sharomi2017optimal} and simulating scenarios \citep{wu2006reducing, carvalho2019mathematical}. However, there are some recent papers considering the use of $R_0$ to analyze and develop epidemic control methods. \citet {ball1986unified} use the average $R_0$ for optimal vaccination policies in a population partitioned into households. \citet{artalejo2013exact} also suggest to use $R_{e0}$ and $R_p$ to design control strategies for prevention of an outbreak. They consider Markov chains while modelling disease spread, so they assume exponential infectious period. $R_{0e}$ indicates the exact number of secondary cases produced by a single infective while $R_p$ denotes the exact number of secondary cases produced by all currently infected individuals until first recovery. Markov chains with exponential infectious period and homogeneous mixing assumption allow them to obtain exact measures and valuable insight rather than an expectation. Since we consider a totally non homogeneous population and a discrete infectious period distribution, we can only measure the expected number of secondary cases  produced by a unique given infected case that changes with the selected initially infected individual and we call it individuals $R_0$.

In this paper, we make three main contributions. Firstly, we introduce individual $R_0$ as the expected number of secondary cases produced by a unique given initially infected individual. Individual $R_0$ is an expectation rather than the exact number of secondary cases so we can propose a general formula for individual $R_0$ in this paper that is applicable to all types of heterogeneous populations with any size. Our second contribution is to present that it is possible to assess intervention strategies by using the exact formula for individual $R_0$ without reverting to simulation. It is possible to assess the impact of intervention strategies by their capability to reduce individual $R_0$ values. Also, a maximum individual $R_0$ value smaller than 1 guarantees that an outbreak is impossible. Lastly, intelligent intervention strategies can be identified based on individual $R_0$ values. We propose a vaccination strategy such that the individual with greatest individual $R_0$ are vaccinated first. In order to choose the individual who is vaccinated next, we recalculate individual $R_0$ values for the unvaccinated individuals and choose the individual with the greatest individual $R_0$ again. Thus, our vaccination strategy is to vaccinate individuals one by one by choosing the susceptible having the largest individual $R_0$.

The paper is structured as follows. In Section 2, the notion of $R_0$ for non homogeneous models is explained and  the general formula to calculate individual $R_0$ values for non homogeneous populations is presented. In Section 3, basic non homogeneous models for influenza spread are presented. Then, in Section 4, we evaluate some intervention methods applied for these models using their individual $R_0$ values. Moreover, we propose intelligent vaccination plans based on individual $R_0$. We also consider a model with overlapping mixing groups and check how individual $R_0$ values changed due to intervention strategies in Section 5. Finally, concluding remarks are given in Section 6. 

\section{The Notion of $R_0$ for Non Homogeneous Models}
\label{The Notion of $R_0$ for Non Homogeneous Models and Individual $R_0$ Calculation}

In this paper, a stochastic SIR model is considered in a non homogeneous population. Stochastic SIR models in large non homogeneous populations grew popular among practitioners in recent years, see eg. \citep{longini2005containing, ajelli2010comparing}. The infection probability between an infected and a susceptible individual is modelled with a comparatively small number of parameters assuming mixing in overlapping mixing groups. The detailed structure of a population is generated such that the mixing groups match in size and age those of real world census data. As mixing groups typically households, neighbourhoods, communities, schools and work places are considered. In several papers it is assumed without any discussion that the only way to assess the behaviour of such models is simulation. This fact attracted our attention and we aim to develop here an approach to assess the behaviour of such models for large populations using a properly defined basic reproduction number $R_0$ that can be calculated easily also for large populations.

As individuals in a non homogeneous population are not identical, $R_0$ for non homogeneous populations depends on the initially infected individual that is chosen. Thus, different $R_0$ values occur for different initially infected individuals. In agent based simulation literature, the value of $R_0$ for the entire non homogeneous population is generally estimated by assuming ``the random case''. Thus, the initially infected individual is selected among the population with equal probability for every individual. Then, they use an average to calculate $R_0$ \citep{longini2005containing}. It is also suggested to estimate age dependent $R_0$ values and calculate overall $R_0$ as a weighted average of age dependent attack rate patterns \citep{germann2006mitigation}. The studies which use branching process methods to calculate $R_0$ also considers $R_0$ for non homogeneous populations as a mean of different secondary cases for different initially infected individuals \citep{ball2002optimal}. In these studies, populations with different mixing levels and moderate size are considered based on census data. However, average $R_0$ is estimated via simulation without exact solution and it cannot be used to assess the possibility of an outbreak anymore.That this is not a sensible approach can be demonstrated with a very simple example:

A population with $N=200$ is composed of two sub-populations of equal size $A$ and $B$. An infective from A infects a susceptible from A with probability 0.003 and a susceptible from B with probability 0.0005 during his total infectious period. Furthermore, an infective from B infects a susceptible from B with probability 0.015 and a susceptible from A with probability 0.0005 during his total infectious period. We can easily calculate: The expected number of secondary cases for a single starting infective of $A$ is 99(0.003)+100(0.0005)=0.347 and for a starting infective of $B$ it is 1.535. Taking the average over all individual we get 0.941. A value of $R_0$ below one should indicate that an outbreak is impossible, but in our little example it is clear that an outbreak in group $B$ is likely if the first infective is of group $B$. And in such a case also several individuals of group $A$ are likely to be infected.

$R_0$ for non homogeneous models is studied especially by using Markov models since they allow to calculate $R_0$ exactly in the literature. However, there are some problems with Markov modelling of disease spread. Markov chain processes requires exponential disease time which is clearly unrealistic. Meanwhile, the complexity of Markov models for non homogeneous populations increases exponentially due to the size of state space, so the exact distribution of $R_0(i)$ can only be calculated for very small populations. It is clear that even for moderate sized populations the state space for such a model is huge. This makes numerical calculations so difficult that \citet{lopez2016stochastic}, who considers a similar but continuous time model with exponential disease times and develops numerical methods to calculate the distribution of important stochastic descriptors, stresses even in the title of the paper that this is only possible for small networks.

Following the simulation literature, we consider a simple discrete time stochastic model. Important is that we allow a very general mixing structure assuming that in a finite population of size $N$ we know all probabilities $p_{ij}$ that within one time-step (in practice typically one day) an infected individual ``$i$'' transmits the disease to a susceptible individual ``$j$''. It turns out that it is also sometimes necessary to allow the possibility that the infection probabilities change with time. In such cases we will write $p_{ijt}$.

The estimation $p_{ij}$ for each pair of $i$ and $j$ is possible for small population sizes like hospitals etc. \citet{laskowski2011agent} implement an agent based modelling for the spread of influenza like disease in an emergency department. They model patients as occupying a circular space with a radius of 60cm and define different contact types like close and casual contacts based on the distance between agents. Moreover, they consider a basic patient flow model throughout which agents come into contact with each other and the probability of infection is found based on the agent distance during contact and the duration of the contact. However, the estimation of $p_{ij}$ is very difficult for large population sizes. The overlapping mixing groups approach is mainly suggested for estimating $p_{ij}$ in large populations. Individual based models for disease spread have been implemented during the last 50 years, but it has been popular recently due to the lack of both data and advanced computational availability in the past \citet{yang2008individual}. \cite{carley2006biowar} propose a scalable city wide multiagent network numerical model where agents are embedded in social, health, and professional networks. The model allows to define heterogeneous population mixing by agent and social networks characteristics based on real data from census, school districts, general social surveys, etc. \cite{bian2004conceptual} presents a conceptual framework for an individual based spatially explicit epidemiological model based on the following assumptions: (1) individuals are different so age groups are needed; (2) an individual has contacts with a finite number of individuals in different clusters like home and workplace; (3) individuals travel between clusters; and (4) the individuals have different contact rates such as fewer contacts for retired individuals than employed individuals. Thus, two types of contacts are defined: those within a group and those between groups. Moreover, the shift from population based models to individual based models is explained by the rapid improvements in computing power and availability of spatial data. \cite{longini2004} compare the efficiency of the use of anti viral drugs and vaccination for a population with 2,000 persons who are stochastically generated by the age distribution and approximate household size published by the US Census Bureau. \cite{yang2008individual} study an individual space time activity model for the target city, Eemnes in the Netherlands based on an activity survey data, a synthesized household data, land use data, and PC6 statistical data.

The agent based models collect the infection data at individual level and become more realistic. However, its increased complexity also brings too much a burden for model structure and requires simulation. Moreover, $p_{ij}$s are required to be computed for individual based models by considering the infection probabilities $p_f$, $p_s$, $p_n$, and $p_c$ in mixing groups of “households”, “school and play groups”, “neighbourhoods” and “communities”, respectively that are also changing with age groups. Then, if infection events between different mixing groups are assumed to be independent, the probability of infection between two individuals $i$ and $j$ during a day is calculated as

\begin{equation}
p_{ij} = 1-(1-p_c)^{I_C(j)}(1-p_n)^{I_N(j)}(1-p_w)^{I_W(j)}(1-p_f)^{I_F(j)}
\label{eomg}
\end{equation}
where the indicator function of a subset is defined as
\begin{align}
I_M(j) = \left\{ \begin{array}{l}
1\begin{array}{*{20}{c}}
{}&{}&{}&{}&{}&{}
\end{array}\begin{array}{*{20}{c}}
{\begin{array}{*{20}{c}}
{}&{}
\end{array}}&{}&{}&{\begin{array}{*{20}{c}}
{}&{}&{j \in Mixing\ Group \ M \  of \ i}
\end{array}}
\end{array}\\
0\begin{array}{*{20}{c}}
{}&{^{\begin{array}{*{20}{c}}
{}&{}
\end{array}}}&&&&&&&&&&&&&&&{otherwise.}
\end{array}
\end{array} \right. \nonumber
\end{align}
That implies that individuals i and j can mix in different mixing groups in set M (community, neighborhood, school, work, family etc.). So the indicator function is one for several j.

The state of our model is described by the state vector holding the state $S$, $I$ or $R$ for all $N$ individuals. In one time step a susceptible individual $j$ is infected by a single infected individual $i$  with probability $p_{ij}$. If there is more than one infected individual the assumption that these infections are independent of each other leads to the total infection probability for individual $j$ :
\[p_j = 1-\prod_{i \in I}\left(1-p_{ij}\right),
\]
where $I$ denotes the set of infected entities. The new infections are thus a sequence of $|S|$ independent Bernoulli trials with probabilities $\{p_j| j\in S\}$, where $S$ denotes the set of all susceptible individuals. To pass from state $I$ to $R$ we use the model assumption that the disease times of all individuals are independent and follow a discrete distribution with probability mass function $f_D(d)$ with
$d=1,2,\dots$ .

For non-homogeneous mixing it is obvious that we need, like suggested in \citet{lopez2016stochastic}, a definition of $R_0$ that considers which individual is the single starting infected. As we consider here large populations, we use the simple classical definition of $R_0$ and define:
\[ R_0(i)= \hbox{E[secondary cases for starting with a unique infected individual $i$]}
\] 
and call it individual $R_0$.

One important advantage of individual $R_0$ is that it can be calculated easily also for large populations. To develop the formula we first have to calculate the probability $\widetilde{p}_{ij}$ that susceptible $j$ is infected by infectious $i$ during the total disease time of $i$.
This is easily done using conditioning on the disease time $D$:
\begin{equation}
\widetilde{p}_{ij}= \sum_{d=0}^{\infty}[f_D(d)(1-(1-p_{ij})^d)].
\label{epij}
\end{equation}
 It is also sometimes possible that the infection probabilities change with time written as $p_{ijt}$. Then, the probability $\widetilde{p}_{ij}$ can be calculated as
\begin{equation}
\widetilde{p}_{ij}= \sum_{d=1}^{\infty}[f_D(d)(1-\Pi_{t=1}^d(1-p_{ijt}))].
\label{epijt}
\end{equation}

Note that also for a disease time distribution with unbounded domain it is not difficult
to calculate a close approximation of  $\widetilde{p}_{ij}$ as the error commited by a cut off of the sum after $d=d_m$ is obviously always smaller than $1-F_D(d_m)$ and can thus be easily controlled.   
Individual $R_0(i)$ is then simply the ''column sum of the matrix $\widetilde{p}_{ij}$'' or more precisely the sum of all $\widetilde{p}_{ij}$'s for $i$ fixed and $j=1,2,\dots,i-1,i+1,i+2,\dots,N$:
\begin{equation}
R_0(i) = \sum_{j:j\neq i }\widetilde{p}_{ij}.
\label{eR0}
\end{equation}

The complexity of  calculating $R_0(i)$ in (\ref{eR0}) for $i=1,2,\dots,n$ is in total $O(d_m\,N^2)$, where $d_m$ denotes the size of the domain of the disease time $D$ for bounded disease time or the cut off value of the infinite sum for the case that $D$ has an unbounded domain.

\subsection{Use of Individual $R_0$ on Intervention Analysis}

A main aim of building agent based simulation models for influenza spread is the assessment of interventions. How is the spread of the disease changed for instance, when 
\begin{itemize}
\item 15 percent of all individuals are vaccinated;
\item anti-viral drugs are given to all members of a household when one member turns out to be infected;
\item when 50 percent of all infected would stay at home after the first day of the disease. 
\end{itemize}

How can the calculation of all $R_0(i)$ values  help to asses the behavior of the disease spread? As we have demonstrated with the help of a simple example above, the average of all $R_0(i)$ values does not allow a direct assessment. But it is easy to see that if $\max_iR_0(i)$ is smaller than one, an outbreak is impossible. If that value is above one the behavior is not certain but an outbreak is possible.

Like for many other interventions also for the second and third intervention example above it is obviously necessary to assume that the $p_{ij}$ values change with time and are denoted by $p_{ijt}$ on day $t$. The $\widetilde{p}_{ij}$'s are obviously calculated using Equation \ref{epijt}. To obtain the $R_0(i)$ we need again the column sums given in (\ref{eR0}).

To quantify the influence of such interventions it is first necessary to decide how the parameters of the model are changed by the intervention. Here it may be necessary to make assumptions (or guesses) how the infection probabilities are changed; if we consider the case that when 50 percent of all infected would stay at home after their first day of infection is an example where it is clear that people staying at home have infection probabilities of zero with all individuals not belonging to their household.

\section{Some Non Homogeneous Population Structures for Influenza Spread}

It is possible to calculate individual $R_0$ values exactly for all non homogeneous models using Equation \ref{eR0}. In this part, we calculate $\widetilde{p}_{ij}$ and individual $R_0$ values for some non homogeneous population structures in the literature that are well applicable for airborne diseases like influenza. We need a discrete disease time for influenza and assume like \citet{longini204} that the probability mass function of 3, 4, 5 and 6 days with probabilities 0.3, 0.4, 0.2, and 0.1 respectively . Moreover, we consider two different non homogeneous population models. Then, in Section 4, we evaluate some intervention methods applied for them.

\subsection{Model with Multiple Cities}

We consider a network of cities around the world connected by transportation. This model is commonly referred as meta population model in the literature suggested by \citet{levins1968evolution}. It includes several sub populations in which perfect mixing is assumed. Individuals travel between the cities leading to disease spread according to probabilistic rules based on the population size and the travel frequency between the cities. Population size and travel data can be obtained from different available sources (e.g. Population Division, U.S. Census Bureau 2004).  

Consider now three cities, numbered 1, 2 and 3.  Assuming symmetry in travel, we consider a function $p_{ij}$ given in Table \ref{probmetapopulation} and compute individual $R_0$ values by applying Equation \ref{eR0}. In big cities, it is standard to assume that $R_0$ is the same in a homogenous population. Thus, the expected number of individuals infected by a single infected individual is not increasing with the size of the population and the probability to meet and potentially to infect another individual is reduced as the population size increases (Lund et al., 2013). Therefore, we assume greater infection probabilities within a city with smaller population sizes. To obtain the infection probabilities between cities is more complicated and challenging and it is beyond the scope of this work \citep{lund2013effects}. Here, we take a simplistic view and assume that the travel frequency is the greatest between city 1 and city 2 and the smallest between city 1 and city 3 by considering the distances between cities and population sizes. Furthermore, the infection probabilities between cities are considered to be around 2 percent of the infection probabilities within the same city. However, it is also possible to obtain travel frequency data for better estimation. Moreover, the reported $R_0$ values for the basic reproduction number in a fully susceptible population is in the range of 1.6 to 2.4 for influenza \citep{germann2006mitigation}. Thus, while setting the infection probabilities, we target to obtain average $R_0$ 1.7 like in the study of \citet{longini2004}. We estimate the infection probabilities by dividing target $R_0=1.7$ over expected disease time and the number of susceptible individuals. We also include some super spreaders in this example supposing there are some individuals who often travel. Because individuals in the same city have identical characteristic, the number of different individual $R_0$ values in this case is equal to the number of cities plus one for the super spreaders. The corresponding individual $R_0$ values are given in the last column of Table \ref{probmetapopulation} computed by using Equation \ref{eR0} consistent with the simulation results.

\begin{table}\footnotesize 
\vskip\baselineskip 
\begin{center}
\begin{tabular}{| c| c| c| c| c|c|c|}
\hline
\shortstack{} &\shortstack{Pop.\\ Size}& {City 1} & {City 2}  &{City 3} & \shortstack{Super\\ Spreaders}& $R_0(i)$\\
\hline
\shortstack{City 1} &{746} & {5.20e-4} & {1.30e-5}  &{8.66e-6} &{1.00e-3} &1.673 \\
\hline
\shortstack{City 2}&500&1.30e-5&7.80e-4&1.04e-5&1.50e-3& 1.714\\
\hline
\shortstack{City 3}&746&8.66e-6&1.04e-5&5.2e-4&1.00e-3&1.668 \\
\hline
\shortstack{Super-spreaders}&8&1.00e-3&1.50e-3&1.00e-3&0&9.174\\
\hline
\end{tabular}  
\caption{Population Sizes and Infection Probabilities for The Population with Multiple Cities.}
\label{probmetapopulation}
\end{center}
\end{table}

\subsection{Model with a Population of Households}

We also consider a population partitioned into several households similar to \citet{ball2002optimal} since the household based public health interventions are important to prevent the spread of infectious diseases. Moreover, the two levels of mixing is also important for the behaviour of the epidemic.

Lets consider that an infected individual infects a household member with probability $p_h$ and other individuals with probability $p_c$. $p_h$ is selected considerably higher than $p_c$ since individuals in the same household have closer contacts. If we denote the family members of individual $i$ as set $N(i)$, the infection probabilities for individual $i$ are
\begin{equation}
  p_{ij}=\begin{cases}
    p_h, & \text{if j $\in$ N(i)}\\
    p_c, & \text{otherwise}.
  \end{cases}
\end{equation} 
We assume that $p_h$ and $p_c$ are 0.0001 and 0.06 respectively for our intervention analysis and we consider 498 households each consisting of four individuals. Further, there might be some individuals in the population who meet with other people more frequently than other individuals eg. due to their work. We call such people super spreaders. In the literature, super spreaders are defined as the individuals infecting more contacts than others. We assume that each infected super spreader infects with probability $p_s=0.0008$ and that there are 8 super spreaders in the population.   

\section{Intervention Analysis}
Intervention methods aim to change the characteristics of the spread of a disease by changing the infection probabilities $p_{ij}$. We suggest to assess  the impact of intervention strategies by calculating and comparing the individual $R_0$ values of the different scenarios. We consider the models described in section 3 where the individuals within the same group are assumed to behave homogeneously. \cite{colizza2008epidemic} define the usual $R_0$ as a function of disease parameters for each group while a subpopulations reproductive number $R_*$  as a function depending on the diffusion rate of individuals among subpopulations. Thus, a group specific basic reproductive number is considered for a deterministic metapopulation system and the epidemic behaviour on both the global scale and the local scale is determined by $R_*$ and $R_0$, respectively. \cite{barthelemy2010fluctuation} consider a stochastic metapopulation model by taking into account both temporal and topological fluctuations. Moreover, individual $R_0$ computed in this section is also a group specific basic reproduction number by considering both infection among the population members of each group and between the members of different groups instead of two different basic reproduction numbers as in the study of \cite{colizza2008epidemic}. However, individual $R_0$ values can be generalized for every non homogeneous population model like individual based models.  Furthermore, we illustrate some numerical results to demonstrate the use of individual $R_0$ for both developing and assessing intervention strategies including vaccination, social distancing and use of antiviral drugs.

\subsection{Intervention by Vaccination}
For the evaluation of vaccine efficacy, it is assumed that vaccination takes place before the infection starts to spread and that all vaccinated individuals develop immunity. Therefore, vaccinated individuals are not considered as susceptible anymore. For the vaccination as an intervention strategy, it is possible to assume random vaccination in which the individuals who are vaccinated are selected randomly with equal probabilities within the population. However, it is better to use the vaccine efficiently to attain herd immunity by vaccinating a smaller number of individuals.

\citet{ball2002optimal} develop optimal vaccination policies for a population with two levels of mixing and consider optimality in terms of the cost of vaccination program including vaccine, administration, and travel. Here, we propose an intelligent vaccination strategy when assuming that the cost of vaccine is considerably larger than the cost of vaccination. In other words, the aim is to obtain for a fixed number of vaccines the greatest reduction for the maximum individual $R_0$ value . In this vaccination strategy, individuals with large individual $R_0$ are vaccinated first because we both eliminate the greatest individual $R_0$ and obtain the greatest total reduction in the other individual $R_0$ values if $p_{ij}$s are symmetric. Therefore, as a next step all individual $R_0$ values must be recalculated and their values are arranged in non increasing order. Then, the individual who is vaccinated is selected from the top of the list and the individual $R_0$ values for unvaccinated individuals are recalculated. Thus, our intelligent vaccination policy is to vaccinate individuals one by one choosing the susceptible having the largest individual $R_0$. By taking the population matrix, $popm$ and the target number of vaccinated individuals, $v_{critial}$ as input parameters, Algorithm \ref{algorithmvaccination} presents the intelligent vaccination strategy.

\begin{algorithm}
\begin{algorithmic}[1]
\STATE Set v=0
\STATE Compute \mbox{$\tilde{p}_{ij}$ using Equation \ref{epij}}
\FOR {$i=1,2,\ldots,N-v$}
       \STATE Compute \mbox{$R_0(i)$ using Equation \ref{eR0}}
\ENDFOR
\STATE Order \mbox{$R_0(i)$ from largest to smallest}
\STATE Remove individual $i$ with the largest $R_0(i)$ from the population matrix
\STATE Set $v=v+1$. \mbox{If $v < v_{critical}$ go to step 3. Otherwise, stop the algorithm.} 
\caption{Intelligent Vaccination Strategy}
\label{algorithmvaccination}
\end{algorithmic}
\end{algorithm}

In the theory of branching process where $m$ is the expected number of children of each individual, $m<1$ implies the ultimate extinction with probability one. If a non homogeneous branching process is considered, $m$ values are different for different individuals \citep{athreya2006branching}. If the maximum $m$ is smaller than one, then the process will be also extinct with probability one. Since we consider non homogeneous populations yielding different individual $R_0$ values, we guarantee that there will be no outbreak  by reducing all individual $R_0$ values below one. Algorithm \ref{algorithmvaccination} for the intelligent vaccination policy is a greedy heuristic for heterogeneous populations and it approximates to the optimal vaccination policy as the heterogeneity level decreases.

If we consider the population with three cities described in Section 3.1, the intelligent vaccination strategy requires to vaccinate individuals from different cities. The simulation results indicate that a significant proportion of the population has been infected with probability 0.662 without vaccination. To guarantee that the infection is going to disappear before involving a significant number of the population by implementing Algorithm \ref{algorithmvaccination}, we observe that individual $R_0$ values in all cities reduce below 1 if 292 individuals from city 1, 200 individuals from city 2, 290 individuals from city 3 and all 8 super spreaders are vaccinated. Therefore, the minimum number of required vaccinated individuals reducing all individual $R_0$ values to under 1 is found to be 790 where there will be no outbreak controlled by computing the final outbreak size through simulation. As 'in the city infection probabilities', $p_{ii}$ are considerably greater than 'between the cities infection probabilities', $p_{ij}$ where $i\neq j$ for a model with multiple cities, intelligent vaccination strategy based on sequential vaccination also gives us the optimal vaccination strategy for reducing all individual $R_0$ values to under one. Let 291 individuals be vaccinated from city 1 instead of 292 individuals, then more than one individual have to be vaccinated from the other cities to decrease individual $R_0$ value of city 1 below one since $p_{11}$ is considerably greater than $p_{12}$ and $p_{13}$. This also holds for city 2 and city 3. However, it does not always yield the optimal strategy. If we consider an individual based model where each individual has its unique $R_0(i)$, we need to decide which individuals are vaccinated in one step by considering all relationships. Even if it is not possible to vaccinate enough people to reach herd immunity intelligent vaccination is still important in order to have the greatest possible reduction of the individual $R_0$ values. Table \ref{vacmetapopulation} shows how individual $R_0$ values change for an increasing number of vaccinated individuals when using the intelligent vaccination strategy.

\begin{table}\footnotesize 
\vskip\baselineskip 
\begin{center}
\begin{tabular}{| c| c| c| c| c|c|c| c|c|c| c|c| c|c|c| c|c|c|c|}
\hline
{} &  \multicolumn{2}{c|}{\shortstack{8 Vaccinated\\ Individuals}} & \multicolumn{2}{c|}{\shortstack{108 Vaccinated\\ Individuals}}&\multicolumn{2}{c|}{\shortstack{208 Vaccinated\\ Individuals}}&\multicolumn{2}{c|}{\shortstack{308 Vaccinated\\ Individuals}}&\multicolumn{2}{c|}{\shortstack{790 Vaccinated\\ Individuals}}\\
\hline
\shortstack{City} & \shortstack{Vacc.\\ Ind.}   & \shortstack{$R_0(i)$}&\shortstack{Vacc.\\ Ind.}& \shortstack{$R_0(i)$}&\shortstack{Vacc.\\ Ind.}& \shortstack{$R_0(i)$}&\shortstack{Vacc.\\ Ind.}& \shortstack{$R_0(i)$}&\shortstack{Vacc.\\ Ind.}& \shortstack{$R_0(i)$}\\
\hline
\shortstack{1}&0 &1.644&35&1.563&73&1.479&111&1.397&292&0.999\\
\hline
\shortstack{2}&0&1.671&32&1.563&56&1.479&81&1.396&200&0.999\\
\hline
\shortstack{3}&0&1.639&33&1.562&71&1.480&108&1.397 &290&0.999\\
\hline
\end{tabular}  
\caption{{Individual $R_0$ Based Intelligent Vaccination with Different Number of Vaccinated Individuals for The Population with Multiple Cities}}
\label {vacmetapopulation}
\end{center}
\end{table}

For the population partitioned into households described in Section 3.2, the individual $R_0$ value for the 1992 individuals living in households is 1.509. The sequence of intelligent vaccination starts with the super spreaders. Then, one individual is vaccinated from every family. To reduce the maximum individual $R_0$ value below 1, vaccination of one individual from every family is not sufficient so the vaccination continues with the vaccination of second individuals from each family. It is easy to calculate that the maximum of individual $R_0$ values drops below one when two individuals are vaccinated in 139 families while only one individual is vaccinated from the remaining 359 families. The two resulting individual $R_0$ values are 0.999 (for 1077 individuals) and 0.777 (for 278 individuals). The minimum number of vaccinated individuals required for herd immunity is thus found to be 645 and. Furthermore, Table \ref{vachousehold} presents the number of susceptibles with their corresponding individual $R_0$ values if 8, 257, 506 and 755 individuals are vaccinated respectively. Table \ref{vachousehold} indicates that individual $R_0$ is 1.483 for unvaccinated individuals if 8 individuals are vaccinated while individual $R_0$ is reduced to 1.159 and 1.381 for 747 and 996 unvaccinated individuals, respectively if 257 individuals are vaccinated. If 506 individuals are vaccinated, individual $R_0$ is reduced to 1.057 for all unvaccinated individuals. Moreover, the last two columns of Table \ref{vachousehold} show that individual $R_0$s are reduced to 0.732 and 0.995 for 498 and 747 unvaccinated individuals if 755 individuals are vaccinated, so vaccinating more than 645 individuals decreases individual $R_0$ much lower than 1.

\begin{table}\footnotesize 
\vskip\baselineskip 
\begin{center}
\begin{tabular}{| c| c| c| c| c|c|c| c|c|c| c|c| c|c|}
\hline
\multicolumn{2}{|c|}{\shortstack{8 Individuals \\ Vaccinated}} & \multicolumn{2}{c|}{\shortstack{In 50\% families\\one member \\ vaccinated}}&\multicolumn{2}{c|}{\shortstack{In all families\\one member\\ vaccinated}}&\multicolumn{2}{c|}{\shortstack{ In 50\% families two \\ and in 50\% families one \\ member vaccinated}}\\
\hline
\shortstack{$R_0(i)$} &\shortstack{ Num. of\\ Indiv.}  &\shortstack{$R_0(i)$}&\shortstack{ Num. of\\ Indiv.}&\shortstack{$R_0(i)$}&\shortstack{ Num. of\\ Indiv.}&\shortstack{$R_0(i)$}&\shortstack{ Num. of\\ Indiv.}\\
\hline
0&8&0&257&0&506&0&755\\
\hline
1.483&1992&1.159&747&1.057&1494&0.732&498\\
\hline
-&-&1.381&996&-&-&0.955&747 \\
\hline
\end{tabular}  
\caption{
{Individual $R_0$ Based Intelligent Vaccination with Different Number of Vaccinated Individuals for The Population Partitioned into Households}}
\label {vachousehold}
\end{center}
\end{table}

If intelligent vaccination is compared to random vaccination, it is observed that the minimum number of required individuals to be vaccinated to reach herd immunity is much higher for random vaccination and its performance under limited vaccination supply is also clearly worse. 

\subsection{Intervention by Social Distancing}
The simplest intervention strategy that can be considered as a method of social distancing is household quarantine. The effectiveness of household quarantine depends on many additional disease parameters like the time between the start of the infection and the start of the symptoms and the compliance rate indicating the percentage of symptomatic influenza cases who remain at home. Household quarantine can be implemented only some time after the infection starts so we assume that it is implemented after the first day of the disease.

We consider the population partitioned into households only since it is not possible to implement social distancing by the nature of a model with multiple cities. To demonstrate the impact of household quarantine, it is assumed that individuals stay at home after the first day of infection with probability 0.5 suggested in the study of \citet{wu2006reducing}. 
\begin{table}\footnotesize 
\vskip\baselineskip 
\begin{center}
\begin{tabular}{| c| c| c| c| c|c|c| c|c|c| c|c| c|c|}
\hline
\multicolumn{2}{|c|}{\shortstack{Without \\ Quarantine}} & \multicolumn{2}{c|}{\shortstack{Quarantine with \\ Compliance Rate 50\%}}& \multicolumn{2}{c|}{\shortstack{Quarantine with \\ Compliance Rate 80\%}}\\
\hline
\shortstack{$R_0(i)$} &\shortstack{ Number of Ind.}  &\shortstack{$R_0(i)$}&\shortstack{ Number of Ind.}&\shortstack{$R_0(i)$}&\shortstack{ Number of Ind.}\\
\hline
1.509&1992&1.191&1992&0.999&1992 \\
\hline
8.084&8&0&8&0&8\\
\hline
\end{tabular}  
\caption{{Quarantine After First Day of Infection for The Population Partitioned into Households}}
\label{quarantinehousehold}
\end{center}
\end{table}
Table \ref {quarantinehousehold} shows the resulting changed individual $R_0$ values. The important point in Table \ref {quarantinehousehold} is the reduction in the individual $R_0$ values of the household members. Therefore, it is possible to decrease individual $R_0$ values by increasing compliance rate. It may be possible to increase the compliance rate if a viable diagnostic support including virological testing is available. Thus, we search the compliance rate to attain herd immunity for the household model. We observe that household quarantine must be accepted by at least $80\%$ of the infected to make an outbreak impossible for the population partitioned into households described in Section 3.2. 

\subsection{Intervention by Use of Antiviral Drugs}
Antiviral drugs can be both of prophylactic and therapeutic importance. The use of antiviral drugs that is evaluated here prevents infection given exposure. Therefore, it is assumed that antiviral drugs reduce the probability of transmission to  others and the probability of being infected given exposure. There are no direct estimates of how much antiviral drug will reduce the probability that an infected individual will develop influenza symptoms compared with an infected person who is not using antiviral drugs but these parameters are inferred from household studies of antiviral drugs in the literature \citep{longini2004}. Therefore, considering that family members of the initially infected individual use antiviral drugs, we check how their individual $R_0$ values change for assuming different reduction factors of anti viral drugs. The results in Table \ref{antiviralmixinggroup} indicate that, as expected, the effectiveness of the use of anti viral drugs is strongly influencing to the reduction capability for infection probabilities. 

\begin{table}\footnotesize 
\vskip\baselineskip 
\begin{center}
\begin{tabular}{| c| c| c| c| c|c|c| c|c|c| c|c| c|c|}
\hline
\multicolumn{2}{|c|}{\shortstack{10\%\\ Reduction}} & \multicolumn{2}{c|}{\shortstack{20\%\\ Reduction}}&\multicolumn{2}{c|}{\shortstack{30\%\\ Reduction}}&\multicolumn{2}{c|}{\shortstack{40\%\\ Reduction}}\\
\hline
\shortstack{$R_0(i)$} &\shortstack{ Num. of\\ Indiv.}  &\shortstack{$R_0(i)$}&\shortstack{ Num. of\\ Indiv.}&\shortstack{$R_0(i)$}&\shortstack{ Num. of\\ Indiv.}&\shortstack{$R_0(i)$}&\shortstack{ Num. of\\ Indiv.}\\
\hline
1.448&1992&1.386&1992&1.323&1992&1.258&1992 \\
\hline
7.942&8&7.797&8&7.649&8&7.498&8\\
\hline
\end{tabular}  
\caption{{Use of Anti Viral Drugs with Different Reduction Factors without Household Quarantine for The Population Partitioned into Households}}
\label{antiviralmixinggroup}
\end{center}
\end{table}

Furthermore, we check how effective is the combination of anti viral drugs and household quarantine. The results are given in Table \ref{antiviralmixinggroup3}. We observe that assuming a compliance rate of 50\% and reduction rate 40\% it is possible to prevent an outbreak by using the combined strategy even if this is not possible when using household quarantine and anti viral drugs alone. 

\begin{table}\footnotesize 
\vskip\baselineskip 
\begin{center}
\begin{tabular}{| c| c| c| c| c|c|c| c|c|c| c|c| c|c|}
\hline
\multicolumn{2}{|c|}{\shortstack{10\%\\ Reduction}} & \multicolumn{2}{c|}{\shortstack{20\%\\ Reduction}}&\multicolumn{2}{c|}{\shortstack{30\%\\ Reduction}}&\multicolumn{2}{c|}{\shortstack{40\%\\ Reduction}}\\
\hline
\shortstack{$R_0(i)$} &\shortstack{ Num. of\\ Indiv.}  &\shortstack{$R_0(i)$}&\shortstack{ Num. of\\ Indiv.}&\shortstack{$R_0(i)$}&\shortstack{ Num. of\\ Indiv.}&\shortstack{$R_0(i)$}&\shortstack{ Num. of\\ Indiv.}\\
\hline
1.130&1992&1.068&1992&1.005&1992&0.940&1992 \\
\hline
5.477&8&5.333&8&5.184&8&5.034&8\\
\hline
\end{tabular}  
\caption{{ Use of Anti Viral Drugs and 50\% Household Quarantine with Different Reduction Factors for The Population Partitioned into Households}}
\label{antiviralmixinggroup3}
\end{center}
\end{table}

\section{A Model with Overlapping Mixing Groups}

The overlapping mixing group model tries to imitate the disease spread in a real world community using census data. It requires only a moderate number of parameters. The average $R_0$ for these models is calculated using simulation in the literature (see \cite{longini2004}). The model uses several mixing groups like “households”, “school and play groups”, “neighborhoods” and “communities” with their respective infection probabilities $p_f$, $p_s$, $p_n$, and $p_c$ changing with age groups to model all infection probabilities $p_{ij}$ that can be calculated by using Equation \ref{eomg}.
 
Moreover, following \cite{longini2004} we also consider asymptomatic cases for the overlapping mixing group case as a feature of influenza in real world. Asymptomatic cases are the infected individuals who do not have symptoms. Their infection probabilities are also considered to be smaller than the ones for symptomatic cases. The implementation of intervention strategies like household quarantine and the use of anti viral drugs is impossible for them due to lack of symptoms. However, the result of vaccination is not influenced by adding asymptomatic cases. To calculate individual $R_0$ for the models with both symptomatic and asymptomatic cases, two $\widetilde{p}_{ij}$ values for both the symptomatic and asymptomatic cases have to be calculated using Equation \ref{epijt}. Then, $R_{0,s}(i)$ and $R_{0,a}(i)$ are calculated using the corresponding $\widetilde{p}_{ij}$ values. The final $R_0(i)$ values are obtained by taking the weighted average of $R_{0,s}(i)$ and $R_{0,a}(i)$.

In the study of \cite{longini2004}, a population of 2000 persons in four identical neighbourhoods is considered. Each individual mixes with people in community, neighborhood, family and play groups. Family sizes differ between one and seven. We have a similar model in the study of \cite{longini2004} but we also added a mixing group work for adults. We constitute a population matrix each row of which includes the ID of community, neighborhood, family, school-work and the age group of an individual similar to the rows of Table \ref{popm}.
\begin{table}[htbp]\footnotesize
\vskip\baselineskip 
\begin{center}
\begin{tabular}{| c|  c| c| c|c| c|c|}
\hline
\shortstack{Individual\\ ID} & \shortstack{Family\\ ID} &\shortstack{Size of\\ Family}  &\shortstack{Neighborhood\\ ID}& \shortstack{Community\\ ID} &\shortstack{Age\\ Group} & \shortstack{School-Work\\ ID} \\
\hline
1& 10&1& 100 &1& 6&9001\\
\hline
2&11&2&100&1&6&9001\\
\hline
3&11&2&100&1&3&3001\\
\hline
\end{tabular}  
\caption{Population matrix for a model with overlapping mixing groups}
\label{popm}
\end{center}
\end{table}
So the number of rows of that population matrix  is 2000. The details of the R code for generating such a population matrix based on census data is available from the authors. A major practical problem for this model is the calibration of the probability of infection within each mixing group. We consider the same infection probabilities as in the study of \cite{longini2004}. As disease duration, 3, 4, 5 and 6 days with probabilities 0.3, 0.4, 0.2, and 0.1 is again assumed. In the study, we assume that an infected person is symptomatic with probability 0.67 and an asymptomatic infection is only half as infectious as a symptomatic infection \citep{longini2004}.
\begin{figure}
\centering\includegraphics[width=12cm]{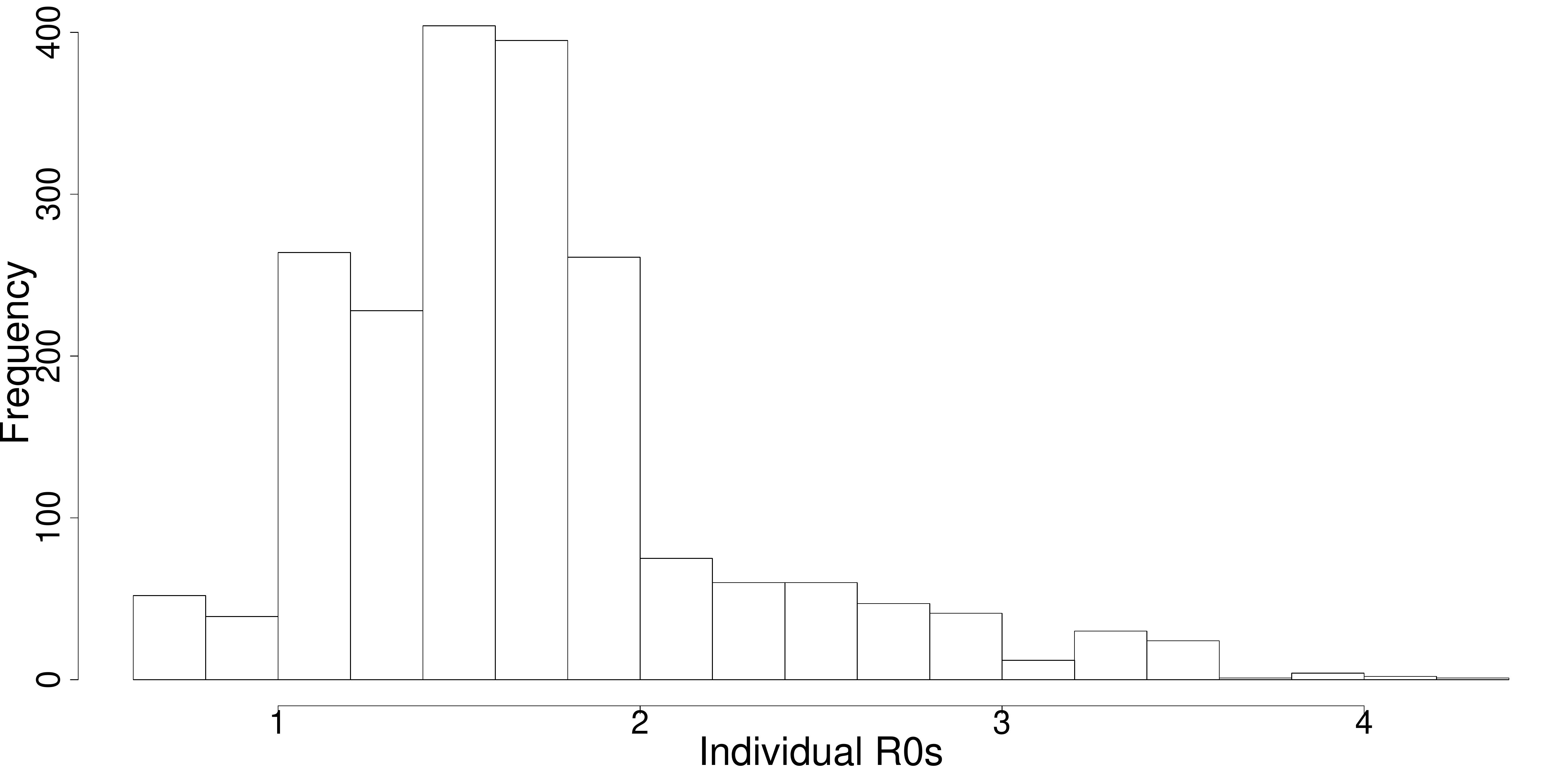}
\caption{The frequency of Individual $R_0$s without household quarantine}
\label{R0}
\end{figure}

In Figure \ref{R0}, we present the histogram of all individual $R_0$ values of the population. The figure indicates that only a small number of individuals in the population have individual $R_0$ values smaller than 1 while most of the population has individual $R_0$ values between 1 and 2. Moreover, \citet{longini2004} estimate $R_0$ as the average of all secondary cases that the randomly selected initial infective person would infect over all mixing groups he belongs to. They empirically calculate $R_0$ and find it as 1.7 with a range of secondary cases from zero to 17. We compute $R_0(i)$ values with Equation \ref{eR0} where $\tilde{p}_{ij}$ is computed by considering the the probability of infection within each mixing group and the population matrix. Moreover, the average of $R_0(i)$ values for $i=1,2,...2000$ is 1.69, so we compute $R_0$ suggested in the study of \cite{longini2004} without implementing simulation. Furthermore, there are many possible $R_0(i)$ values giving the same average and the importance of $R_0(i)$ values increase as the heterogeneity level increases.

In studies with overlapping mixing groups, we found only the suggestion of random vaccination and of random vaccination of children \citep{longini2004, germann2006mitigation}. In these studies, the results of vaccination are analysed estimating attack rates via simulation. However, we suggest to assess the intervention strategies without simulation also for individual based models. Since we can compute individual $R_0$ value for each member of the population exactly, see the histogram in Figure 1, we can compare the frequency histograms of individual $R_0$ values after different intervention strategies are implemented. Moreover, simulation is not needed while vaccinating individuals based on their individual $R_0$ values by implementing Algorithm \ref{algorithmvaccination}. In this section, we apply to simulation only for random vaccination in order to compare intelligent vaccination strategy with random vaccination where the vaccinated individual is selected randomly. We compare the performance of intelligent vaccination and random vaccination by considering 30\%, 50\% and 80\% of the population vaccinated respectively. Moreover, we record the maximum individual $R_0$ value for the intelligent vaccination. However, we record the minimum, median and maximum of maximum individual $R_0$ values for random vaccination because each run yields a different  maximum individual $R_0$ value. In Table \ref{comparisonofvac}, we present the results.

 \begin{table}[htbp]\footnotesize
\vskip\baselineskip 
\begin{center}
\begin{tabular}{| c|  c| c| c|c|}
\hline
\shortstack{Vaccination \\ Percentage} &\shortstack{Minimum of Max\\Individual $R_0$s in\\1000 repetitions} & \shortstack{Median of Max\\Individual $R_0$s in\\1000 repetitions}&\shortstack{Maximum of Max\\Individual $R_0$s in\\1000 repetitions} &\shortstack{Max $R_0$ after\\ Individual $R_0$ Based\\ Vaccination}\\
\hline
0& 4.27 &4.27&4.27& 4.27\\
\hline
30&2.40&3.08&3.73&1.32\\
\hline
50&1.70&2.23&2.98&0.91\\
\hline
80&0.60&0.96&1.54&0.48\\
\hline
\end{tabular}  
\caption{Maximum  Individual $R_0$ Values after Random Vaccination and Vaccination Based on Individual $R_0$ without Household Quarantine}
\label{comparisonofvac}
\end{center}
\end{table}

\begin{figure}
\centering\includegraphics[width=12cm]{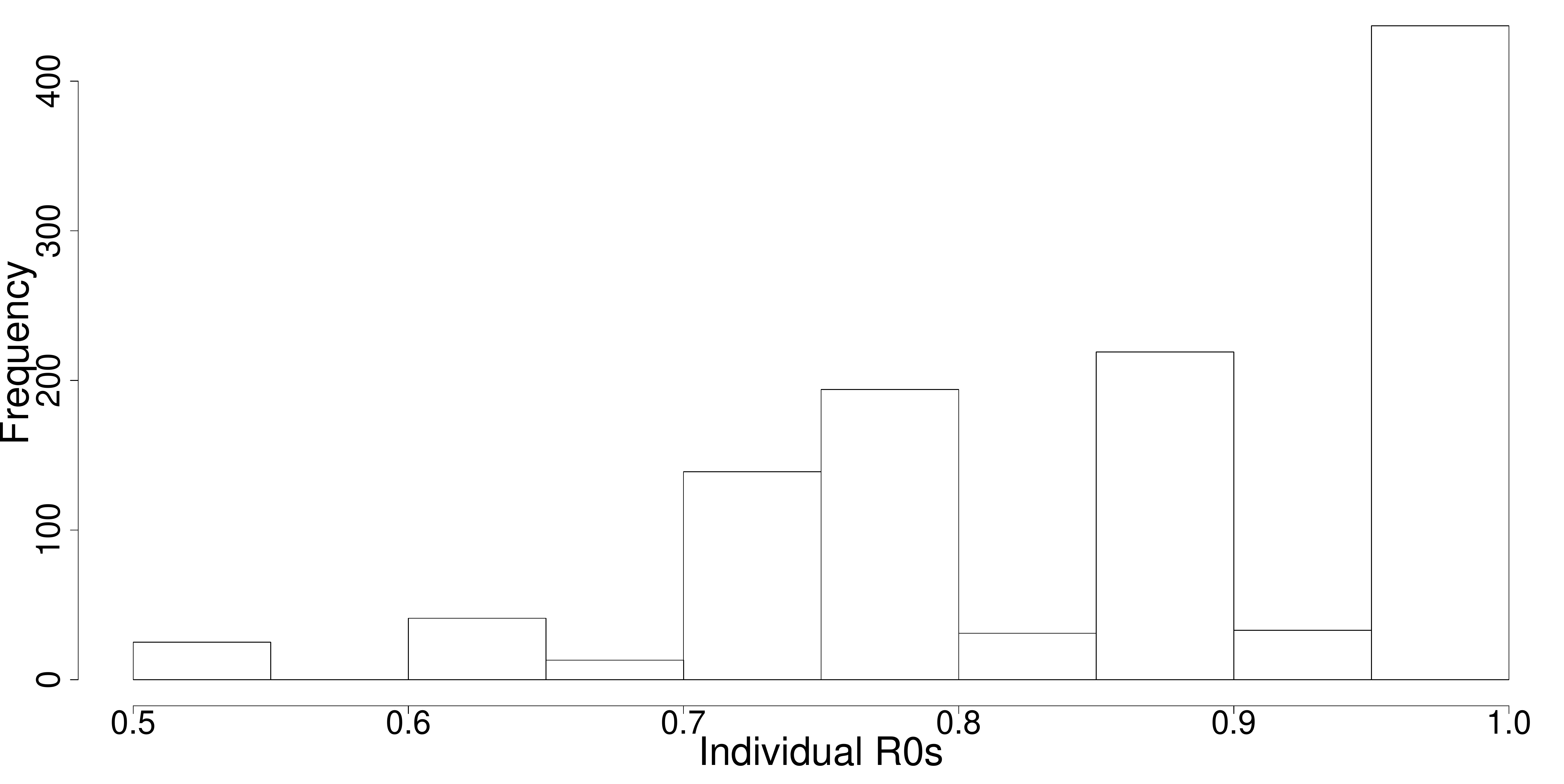}
\caption{The frequency of individual $R_0$s after vaccination without household quarantine}
\label{vacR0}
\end{figure}

Table \ref{comparisonofvac} indicates that 50\% random vaccination of the population cannot reduce maximum individual $R_0$ below 1 in 1000 repetitions while 50\% vaccination based on individual $R_0$ values of the population reduces maximum individual $R_0$ much lower than 1. Thus, similar to the model with multiple cities and the model with a population partitioned into households, we take the advantages of our $R_0$ formula. Furthermore, the minimum required number of vaccinated individuals to guarantee herd immunity is 869. In Figure \ref{vacR0}, we present individual $R_0$ values of the unvaccinated population after vaccination of 869 individuals based on their individual $R_0$s.

\begin{figure}
\centering\includegraphics[width=12cm]{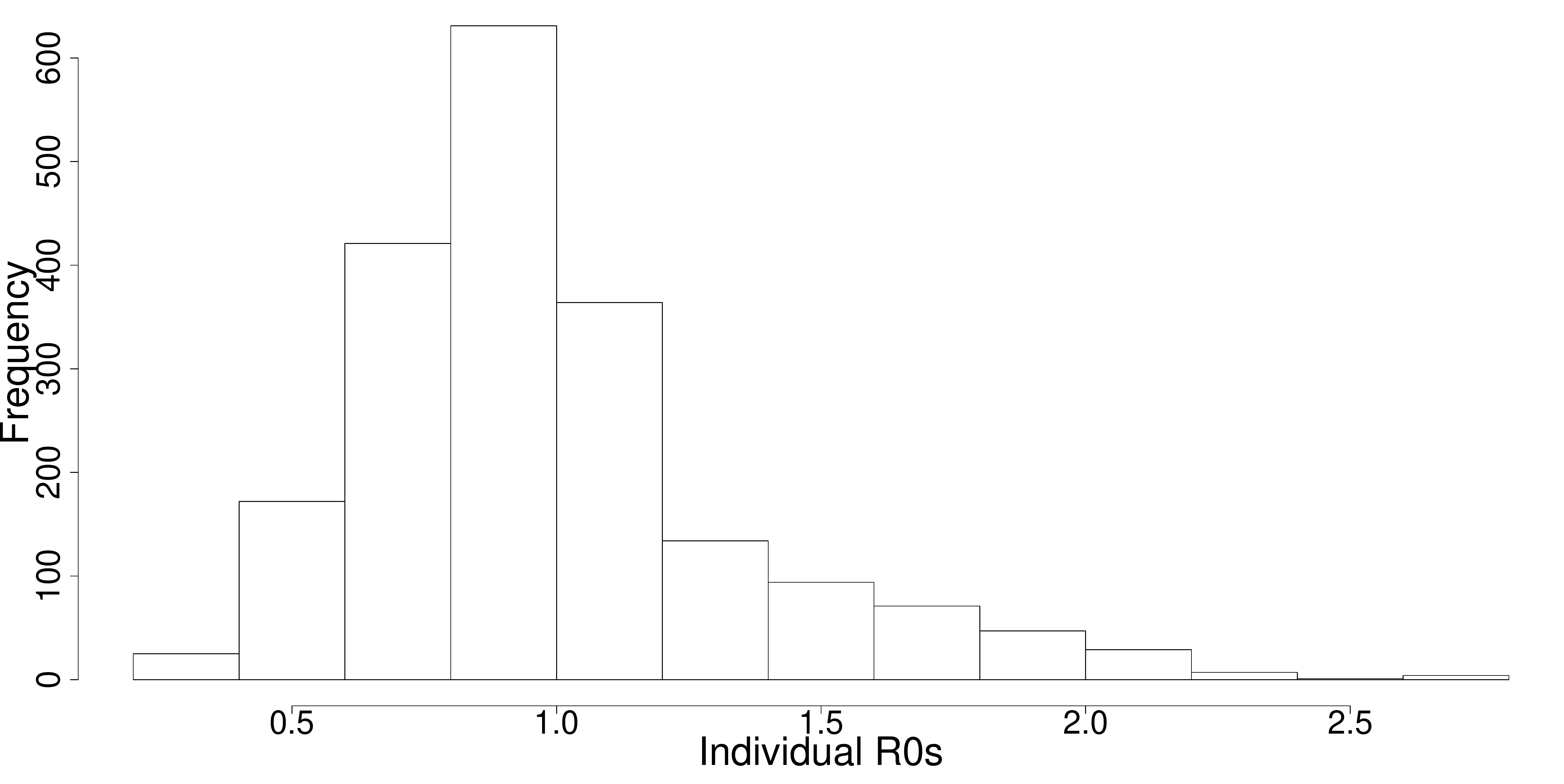}
\caption{The frequency of individual $R_0$s after 80\% household quarantine}
\label{qR0}
\end{figure}

Finally, we also check how individual $R_0$ values change if 80\% of the symptomatic cases stay home after their first day of infection. Figure \ref{qR0} shows that even if a considerable reduction in individual $R_0$ values is obtained, herd immunity cannot be guaranteed since one third of the infectious cases are considered to be asymptomatic and household quarantine cannot be implemented for asymptomatic cases. This is also true for 100\% compliance rate since the actual compliance rate can be at most 67\% that is the percentage of symptomatic cases.

The results certainly depend on disease parameters and population structures but we expect that recursive individual $R_0$ based vaccination gives consistently a better performance. So we can see that the calculation of individual $R_0$ can be a useful tool to assess the performance of vaccination strategies and also to develop vaccination strategies for stochastic models with arbitrary heterogeneous contact structures.

\section{Discussion}

In this paper, we consider a discrete time stochastic SIR model for non homogeneous populations and make three main contributions. Firstly, we introduce individual $R_0$ and propose a general formula for it that is applicable to all types of heterogeneous populations with any size. Our other major contribution is the assessment of intervention strategies by using the formula for individual $R_0$ without reverting simulation. Lastly, we define intelligent intervention strategies based on individual $R_0$ values.

As we have studied the notion of $R_0$ for non homogeneous populations, we introduced individual $R_0$ as the expected number of secondary cases produced by a unique given initially infected individual. In the literature, $R_0$ for non homogeneous populations is either calculated by using Markov chains assuming exponential disease time and small population size or estimated via simulation. Here, we propose a general formula for exact calculation of individual $R_0$ that is applicable to an arbitrary mixing structure and large population size. 

Furthermore, the evaluation of intervention strategies is of practical importance. The effectiveness of these strategies is evaluated by simulation studies comparing the average attack rates or similar characteristics. However, we show that it is possible to assess the impact of the intervention strategies by using directly the individual $R_0$ formula. We analyze the effectiveness of different intervention strategies by their ability to decrease the maximum individual $R_0$ value below one. This method is more accurate than descriptive simulation results to decide how to make an outbreak impossible. However, it is only possible to evaluate strategies that are implemented before infection and immediately after one case is infected by using the individual $R_0$ values of implementing vaccination, household quarantine or the use of antiviral drugs. 

Finally, an intelligent vaccination policy is developed based on individual $R_0$ values. Here, the aim is to obtain the greatest reduction in the maximum individual $R_0$ value for a fixed number of vaccines. It is observed that the number of required individuals to be vaccinated for herd immunity is much higher for random vaccination than vaccination based on individual $R_0$.


\end{document}